\documentclass{sig-alternate-2013}
\usepackage{url}
\usepackage{multirow}
\newtheorem{theorem}{Theorem}[section]
\newtheorem{lemma}[theorem]{Lemma}

\newtheorem{corollary}[theorem]{Corollary}
\usepackage{array}
\usepackage{rotating}
\usepackage{algorithm}
 \usepackage[noend]{algpseudocode}
\usepackage{color}
\usepackage{psfrag}
\usepackage{makecell}

\newcommand{\todo}[1]{\textcolor{red}{TODO: #1}\\}

\newfont{\mycrnotice}{ptmr8t at 7pt}
\newfont{\myconfname}{ptmri8t at 7pt}
\let\crnotice\mycrnotice%
\let\confname\myconfname%

\permission{Permission to make digital or hard copies of all or part of this work for personal or classroom use is granted without fee
provided that copies are not made or distributed for profit or commercial advantage and that copies bear this notice and the full citation
on the first page. Copyrights for components of this work owned by others than ACM must be honored. Abstracting with credit is permitted. To
copy otherwise, or republish, to post on servers or to redistribute to lists, requires prior specific permission and/or a fee. Request
permissions from Permissions@acm.org.}
\conferenceinfo{KDD'14,}{August 24--27, 2014, New York, NY, USA.}
\copyrightetc{Copyright 2014 ACM \the\acmcopyr}
\crdata{978-1-4503-2956-9/14/08\ ...\$15.00.\\
http://dx.doi.org/10.1145/2623330.2623707}

\begin{document}
%
%

\title{On the Permanence of Vertices in Network Communities}

\numberofauthors{1} 
\author{\alignauthor Tanmoy Chakraborty$^{1,a}$, Sriram Srinivasan$^{2,b}$, Niloy
Ganguly$^{3,a}$, \\Animesh Mukherjee$^{4,a}$, Sanjukta Bhowmick$^{5,b}$ \\
\affaddr{$^a$Dept. of Computer Science \& Engg., Indian Institute of Technology, Kharagpur, India -- 721302} \\ \affaddr{ $^b$Dept. of
Computer Science, University of Nebraska, Omaha, Nebraska 68182} \\
\email{\{$^1$its\_tanmoy,$^3$niloy,$^4$animeshm\}@cse.iitkgp.ernet.in\\
\{$^2$sriramsrinivas,$^5$sbhowmick\}@unomaha.edu}
}

\maketitle
\begin{abstract}
Despite the prevalence of community detection algorithms, relatively less work has been done on understanding whether a network is indeed
modular and how resilient the community structure is under perturbations. To address this issue, we propose a new vertex-based metric called
{\it permanence}, that can quantitatively give an estimate of the community-like structure of the network.

The central idea of permanence is based on the observation that the strength of membership of  a vertex to a community depends upon the 
following two factors: (i) the distribution of external connectivity of the vertex to individual communities and {\em not} the total
external connectivity, and (ii) the  strength of its internal connectivity and {\em not} just the total internal edges.

In this paper, we demonstrate that  compared to other metrics, permanence provides (i) a more accurate estimate of a derived community
structure to the ground-truth community and (ii) is more sensitive to perturbations in the network.  As a by-product of this study, we have
also developed a community detection algorithm based on maximizing permanence. For a modular network structure,  the results of our
algorithm match well with ground-truth communities.
\end{abstract}




\section{Introduction}

Finding accurate community structures, i.e., groups of vertices that have more connections within a group than across the groups is one of
the central problems in network analysis. 
Several community detection algorithms have been proposed over the last decade; these algorithms are generally based 
on optimizing certain scoring functions (for example, modularity~\cite{Newman:2006} or conductance~\cite{cond_09}). The output of these
algorithms is an assignment of the vertices to their respective communities, for which the designated parameters are optimal (or
nearly-optimal)\footnote{\scriptsize{In this paper, we consider only the non-overlapping communities.}}.  However, almost all these
detection techniques will always output a set of communities, irrespective of whether the network has an inherent community structure or
not. Moreover, the optimal values of the scoring functions do not provide any insight as to whether a network actually possesses community
structure or not. For example, the highest modularity in the Jazz network is 0.45~\cite{chakraborty} and that of the Western  USA  power
grid is 0.98~\cite{chakraborty, Karrer}. However, it has been observed  \cite{chakraborty, Karrer},
that Jazz has a much stronger community structure than the power grid. 

The key reason for this is that optimization metric such as modularity frequently enforces the detection algorithm to make a choice by arbitrarily breaking ties. While this indeed increases the value of the metric, each such tie-breaking obfuscates the possibility of other community assignments. In grid-like networks, where choices can occur frequently, such tie-breaking can produce inaccurate or insignificant communities, while producing a high scoring function. Although, methods for finding consensus communities \cite{Santo} can indicate whether the communities are significant or not, these techniques are dependent on the number of algorithms used to find the consensus. 

In this paper, we propose a  novel  vertex-based scoring function called {\em permanence} whose optimization encounters much fewer tie-breaking situations in a network. The key idea behind formulating permanence is as follows. Most optimization metrics consider either the degree of a vertex in a community or  the total number of external connections of the vertex (i.e., those connections that are attached with the other neighbors of the vertex outside the community). We posit that the distribution of  the external connections of a vertex is equally important. In particular, our vertex assignment decisions are based not on the total number of external connections but on the maximum number of external connections to any single neighboring community.  To the best of our knowledge, we are the first to make this distinction between the total external connections and their distribution.
Permanence of a vertex thus  quantifies its propensity  to remain in its assigned community and the extent to which it is ``pulled'' \cite{chakraborty} by the neighboring communities. 

The value of  permanence  ranges from 1 (vertex is strongly connected to its assigned community) to -1 (vertex is weakly connected to its
assigned community, and possibly wrongly assigned). If the permanence is zero, this indicates that the vertex is pulled equally by all its
neighbors, all of which are in different communities. The ``pull'' in the metric is modeled as the maximum number of external connections to
any single neighboring community. The introduction of  pull in the formulation  significantly reduces the frequency of tie-breaking
situations that the algorithm has to encounter. In case the ``pull'' from all the neighboring communities is equal for a vertex, we assign
it to a {\em singleton} community (i.e., community containing only one vertex), rather than assigning it to one of the (larger size)
neighboring communities.  

The sum of the  permanence of all vertices, normalized by the number of vertices, provides the permanence of the network. It indicates to what extent, on an average, the vertices of a network are bound to their communities. As with permanence, this value also ranges from 1 to (nearly) -1. Maximizing permanence can be therefore used as an alternative method for identifying communities which constitutes a by-product of the current study. This approach of combining the microscopic (vertex-level) information to obtain the mesoscopic (community-level) information provides a more fine-grained view of the modular structure of the network. As the community structure of the network degrades, so does the value of permanence of the entire network. 

The principal benefits of our approach are -- {\bf (i)} permanence provides a quantitative estimate of the inherent community structure of
the network (Section \ref{permanence_def}),{\bf  (ii)} permanence is comparable (and sometimes better) than several other popular community
scoring functions in identifying good communities (Section~\ref{goodness}), {\bf (iii)} permanence is very sensitive to the different
perturbations of the network -- a  desirable property for a community scoring metric (Section~\ref{parturbation}), {\bf (iv)} for modular
networks, maximizing permanence algorithm is more successful in finding ground-truth communities as compared to several other community
detection algorithms (Section~\ref{comm}), {\bf (v)} community detection using maximizing permanence can reduce the effect of resolution
limit, degeneracy of solutions and asymptotic growth of the optimal value with network size (Section~\ref{comm}).

\section {Definition of Permanence}\label{permanence_def}

In this section, we explain the concepts that lead to the formulation of permanence followed by a definition of the formula. 

\subsection{ Distribution of External Connections}  
In contrast to most optimization metrics that either consider the degree of the vertex in a community or the total number of external neighbors of the vertex, in permanence we consider the distribution of external connections of the vertex to its neighboring communities. A vertex that has equal number of connections to all its external communities (e.g., a vertex with total 6 external connections with 2 to each of 3 neighboring communities) has equal ``pull" from each community whereas a vertex with more external connections to one  particular community  (e.g., a vertex with total 6 external connections with 1 connection each to two neighboring communities and 4 connections to the third neighboring community), will experience more ``pull" from that community due to large number of external connections to it.

This property is demonstrated by a toy example in Figure~\ref{examplefig}(a). If the edge $(a,b)$ is deleted and the edge $(a,c)$ is added, then the number of external connections remains the same, 
and the value of modularity, conductance and cut-ratio are also the same for this change. However, in the initial graph, vertex $a$ had more
``pull" from the community of $b$, in fact proportional to the number of its internal connections, whereas in the modified version $a$ has
equal pull from both the communities of $b$ and $c$. Our  permanence formula, defined in Section~\ref{sec:perm}, takes this distinction
into account. 

Figure~\ref{examplefig}(b) shows a histogram of the fraction of
vertices versus the ratio between the number of total ($E_{sum}$) and maximum ($E_{max}$) external connections for two representative
networks. We notice that -- (i) very few vertices have (closely) similar values of $E_{sum}$ and $E_{max}$ (i.e., ratio=1); the majority 
have
significantly different $E_{sum}$ and $E_{max}$ (ii) the ratio between these two quantities  is not constant;  it is spread over
a wide range of values.  Therefore, we cannot estimate  $E_{max}$ from the value of $E_{sum}$. Consequently,  metrics that are  based on
total number of external connections lack the information as to what extent a vertex may be ``pulled'' by the neighboring communities which
can better estimated by $E_{max}$.  Using $E_{sum}$ can potentially result in frequent ties that need to be arbitrarily resolved by the community detection algorithms based on such metrics. 

{\em When computing permanence, we use the maximum number of external connections, i.e., the maximum pull, to any one external community,
instead of combining all the external connections.}

\subsection{ Strength of Internal Connections} 
\begin{figure}[!ht]
\begin{center}
 \begin{tabular}{c c}
 \multicolumn{2}{c}{\includegraphics[scale=0.2]{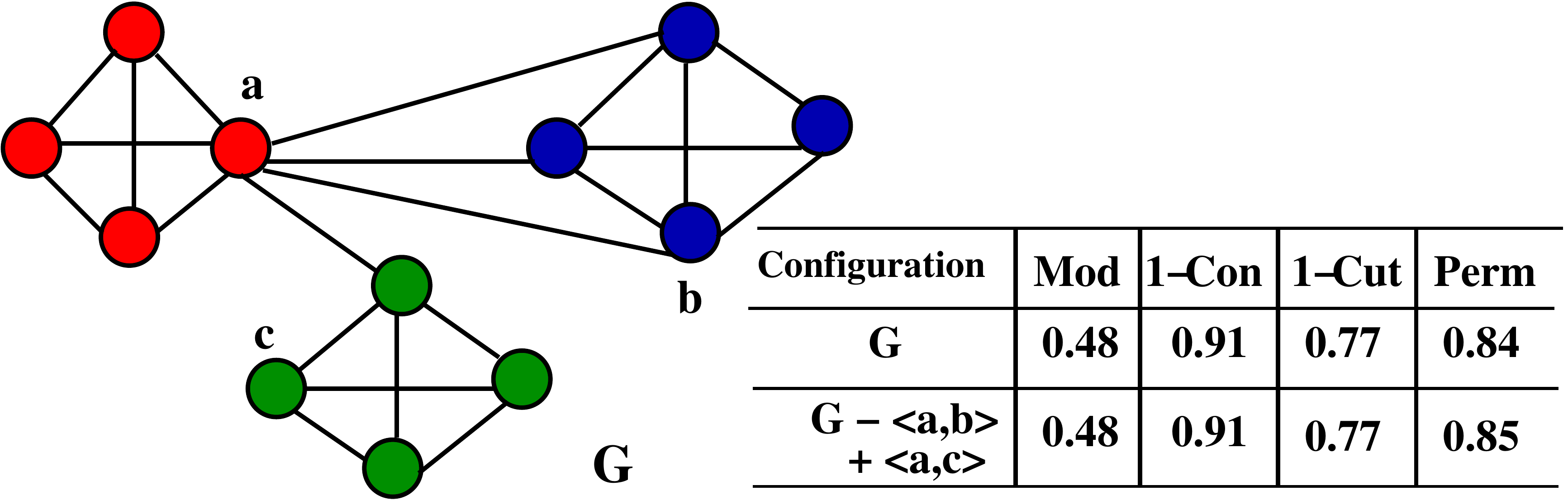}}  \\
  \multicolumn{2}{l}{\scriptsize{(a) Example demonstrating the importance of the distribution of} }\\
  \multicolumn{2}{l}{\scriptsize{external connections.} }\\
  \multicolumn{2}{c}{\includegraphics[scale=0.20]{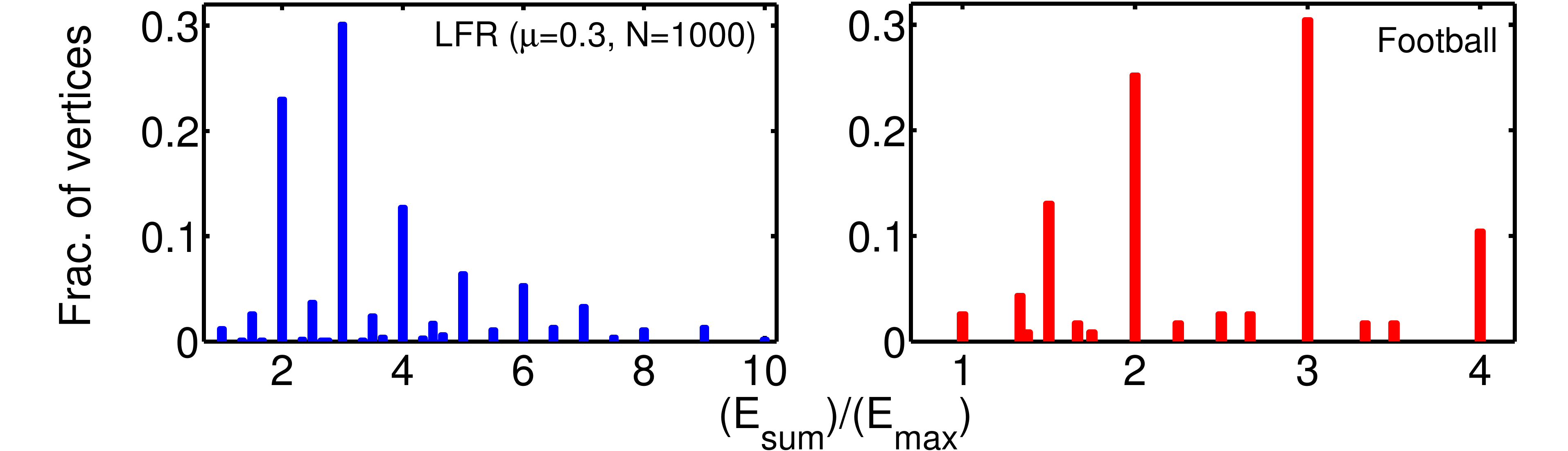}}  \\
  \multicolumn{2}{l}{\scriptsize{(b) Fraction of vertices versus the ratio between the number of }}\\
  \multicolumn{2}{l}{\scriptsize{ total ($E_{sum}$) and maximum ($E_{max}$) external connections.}}\\
      \multicolumn{2}{c}{\includegraphics[scale=0.19]{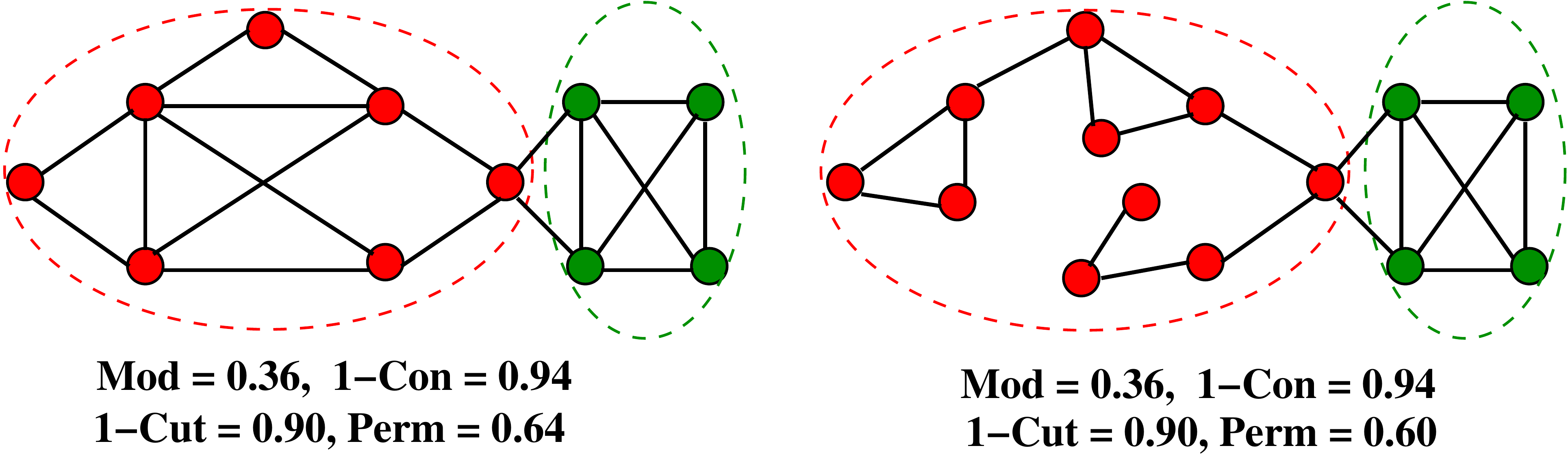} } \\
  \multicolumn{2}{l}{\scriptsize{(c) Two networks with same modularity, conductance and cut-ratio,}}\\
   \multicolumn{2}{l}{\scriptsize{but the left one has more prominent community structure.} }\\
    \multicolumn{2}{c}{\includegraphics[scale=0.18]{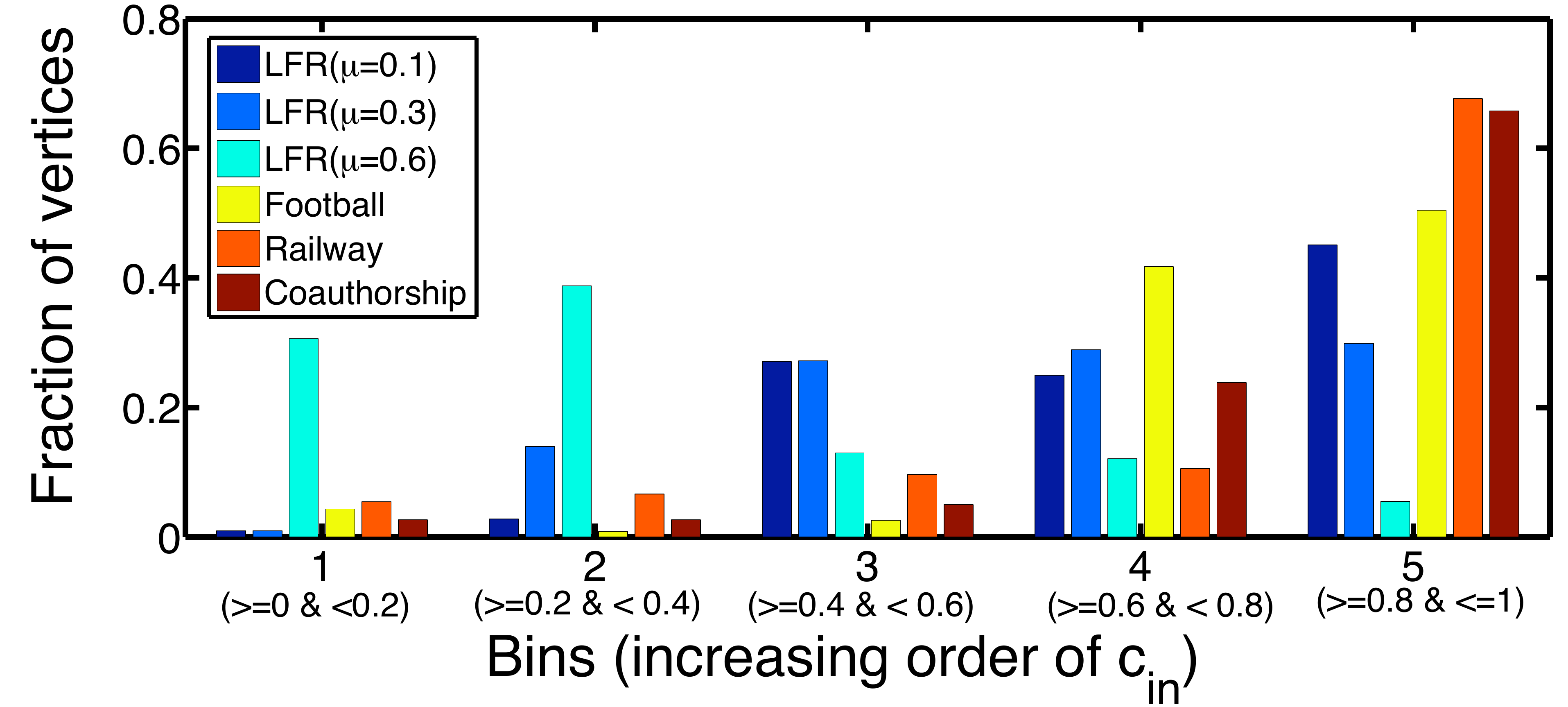}}  \\
  \multicolumn{2}{l}{\scriptsize{(d) Fraction of vertices with a specific range of internal clustering} }\\
  \multicolumn{2}{l}{\scriptsize{ coefficient ($c_{in}$) in LFR and real-world networks.} }\\\\

 \end{tabular}
\end{center}
\vspace{-6mm}
 \caption{(Color online) Toy examples and empirical observations demonstrating the drawbacks of existing community scoring metrics. The communities in
(a) and (c) are distinguished by different colors. }\label{examplefig}
\end{figure}

The internal connections of a community are generally considered together as a whole.
However, how strongly a vertex is connected to its internal neighbors can differ. The toy example of Figure~\ref{examplefig}(c) shows two
networks each having two communities. Both the networks have the same number of edges; and the modularity, conductance and cut-ratio
for the two divisions are exactly the same. However, the vertices on the left-hand graph are more tightly connected to each other than the
vertices on the right-hand graph. To measure this internal connectedness of a vertex, one can compute the clustering coefficient of the
vertex with respect to its internal neighbors. The higher this internal clustering coefficient, the more tightly the vertex is connected to
its community.

As an empirical study, we further obtain the internal clustering coefficient per vertex of the benchmark networks for their ground-truth
communities. Figure~\ref{examplefig}(d) shows a histogram of the internal clustering coefficient versus the number of vertices corresponding
to a specific range of internal clustering coefficient. As can be seen from the histogram, for most vertices the internal clustering
coefficients are generally towards the high range. However, for LFR ($\mu$=0.6) there is a reverse trend. In this network,
there are more vertices with lower internal clustering coefficient. This network by construction has a weaker community structure than the
other networks in the set, and thus quite a few of its vertices are loosely connected internally (see more in Section~\ref{comm}).

{\em To represent whether vertices are tightly connected within their communities, we include the internal clustering coefficient  as a factor in computing permanence.}

\begin{figure}[!h]
\centering
 \includegraphics[scale=0.23]{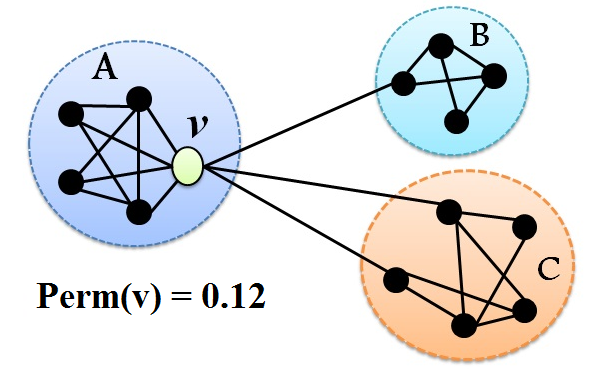}
 \caption{Toy example depicting {\em permanence} of a vertex $v$ (here $I(v)$=4, $D(v)$=7, $E_{max}(v)$=2,
$c_{in}(v)$=$\frac{5}{6}$).}\label{example}
\end{figure}

\subsection{Formulation of Permanence}\label{sec:perm}
Based on our observations  on the distribution of external connections and the internal clustering coefficient, we formulate permanence of a vertex based on the following two criteria that measure the possibility of the vertex remaining in its own community: 

{\bf (i)} The internal connections, $I(.)$, of the vertex $v$ should be more than the maximum connections to a single external community, $E_{max}(.)$, which results more internal pull than the maximum external pull. This criteria is represented in the permanence computation as the ratio of $I(v)$ and $E_{max}(v)$ (indicated by F1 in
Equation~\ref{perm}). If the vertex has no external connections, F1 is just the value of the internal connections. We normalize this value
by the total degree of the vertex, $D(v)$ (indicated by F2 in Equation~\ref{perm}), which ensures that the product of F1 and F2 will be
between 0 (no internal connections) and 1 (no external connections).

{\bf (ii)} Within a specific community, the internal neighbors of the vertex $v$ should be highly connected among each other (i.e., its internal clustering coefficient\footnote{{\scriptsize Note that, internal clustering
coefficient of $v$ is obtained by considering the ratio of the existing connections and the total number of possible connections among the
{\em internal neighbors} of $v$.}}, $c_{in}(v)$, should be high). This criteria
emphasizes that a vertex is likely to be within a community if it is part of a near-clique substructure. For computing $c_{in}(v)$, we assume that each community should have at least three vertices and three internal connections; otherwise, $c_{in}(v)$ is set to 0. When computing permanence, we impose a penalty
based on low internal clustering coefficient (indicated by F3 in Equation~\ref{perm}). The less the internal clustering coefficient, the
more the penalty imposed to the final outcome of the community score. This value also ranges from 0 (no penalty) to 1 (maximum penalty).


We aggregate these two criteria to formulate permanence of a vertex $v$ as follows: 
\begin{equation}\label{perm}
Perm(v)= \Big[  \underbrace{\frac{I(v)}{E_{max}(v)}}_\text{F1}\times  \underbrace{\frac{1}{D(v)}}_\text{F2}\Big] - \Big[ \underbrace{1-c_{in}(v)}_\text{F3}\Big]
\end{equation}
Figure~\ref{example} depicts a toy example for measuring permanence of a vertex $v$. Note that, this formula actually differentiates between the two cases in Figure~\ref{examplefig}(a) with higher permanence value for the case where the external pull is uniform. Similarly, the formula differentiates between the two cases in Figure~\ref{examplefig}(c) by imposing more penalty on the network that has a less tightly knit internal substructure.

\subsection {Boundary Conditions of Permanence }
For vertices that do not
have any external connections, $Perm(v)$ is considered to be equal to the internal clustering coefficient (i.e., $Perm(v) = c_{in}(v)$). The maximum value of $Perm(v)$ is 1 and is obtained when vertex $v$ is an internal node and part of a clique. The lower bound of $Perm(v)$ is  close
to -1. This is obtained when $I(v) \ll D(v)$, such that $\frac{I(v)}{D(v)E_{max}(v)} \approx 0$ and  $ c_{in}(v)=0$. Therefore for every
vertex $v$, $-1 < Perm(v) \leq 1$. The permanence of a graph $G(V,E)$, where $V$ is the set of vertices and $E$ is the set of edges,
is given by $Perm(G)=\frac{1}{|V|}\sum_{v \in V}Perm(v)$. For a graph $G(V,E)$, the range is $-1 < Perm(G) \le 1$. 

$Perm(G)$ will be closer to 1 as more vertices have high permanence, that is more vertices are in well-defined communities. This can happen only if the network has a strong community structure. 
The maximum value obtained is when $G$ consists of a series of disconnected cliques. If there is a vertex bridging between two cliques, then the highest overall permanence will be obtained if each clique acts as a separate community and bridging vertex forms a singleton community. For a grid, the best value of $Perm(G)$ will be zero, i.e., each vertex is assigned to a singleton community. 

\if{0}
\subsection{Advantages of Permanence over Modularity } Permanence has two significant advantages over modularity -- (i) the value of
permanence is not as much affected by the symmetric growth of
the network size, and (ii) the value tends to fall sharply as the community structure deteriorates. Table~\ref{change} supports both these 
points. If we increase the size of the LFR network (see in Section \ref{dataset}) keeping $\mu$ (average ratio between the external
connections of a node to its degree) constant, the change in permanence is much less compared to the change in modularity (details in
Section \ref{permanence_property}, Table~\ref{asymptotic}). On the other hand, given a fixed network size,
if we worsen the quality of the community structure by increasing $\mu$, the value of permanence decreases significantly. The negative value
of permanence clearly indicates the less stringent community structure in LFR ($\mu$=0.6) network. Therefore, the value of permanence is a strong indicator of the quality of the community structure of a network, and  this value is very sensitive to changes
 in the community structure.    

\vspace{-4mm}
\begin{table}[!h]
\caption{Change in modularity and permanence with the increase of $\mu$ and $n$ in LFR network.}\label{change}
\begin{center}
\scalebox{0.8}{
  \begin{tabular}{|c|c|c|c|c|c|c|}
  \hline
  & \multicolumn{3}{c|}{Modularity} & \multicolumn{3}{c|}{Permanence} \\\hline
 \diaghead{\theadfont $NNN\mu$} {$\mu$}{n}      & 1000 & 3000 & 6000 & 1000 & 3000 & 6000\\\hline
 0.1   & 0.85 & 0.88 & 0.89 & 0.56 & 0.57 & 0.57 \\\hline
 0.3   & 0.66 & 0.68 & 0.69 & 0.26 & 0.27 & 0.27 \\\hline
 0.6   & 0.46 & 0.48 & 0.49 & -0.13 & -0.13 & -0.13 \\\hline
 \end{tabular}}
 \end{center}
\end{table}

\fi

\vspace{-2mm}
\section{Experimental Setup}\label{dataset}
In this section, we provide a brief overview of the datasets, metrics and comparative methods that we use for our experiments.

\subsection{Test Suite of Networks}  
We use the {\bf LFR benchmark} model~\cite{Lancichinetti} to generate synthetic networks along with their ground-truth communities. Users can
 specify the following properties: number of nodes ($n$), average ($<k>$) and maximum ($k_{max}$) degree, the degree distribution,  the community size distribution, and the
mixing-coefficient ($\mu$). The mixing coefficient represents the ratio (in average) between the external connections of a node to its degree. Thus the lower the value of $\mu$, the stronger the community in the network. For our experiments, we set the number of nodes as 1000, and $\mu$ as 0.1, 0.3 and 0.6 (unless mentioned otherwise). For the rest of the parameters, we use the default values mentioned in the authors'
implementation\footnote{{\scriptsize\url{https://sites.google.com/site/santofortunato/inthepress2}}} \cite{Lancichinetti}.

We also  use three real-world networks\footnote{{\scriptsize All the datasets are publicly available at
\url{http://cnerg.org/permanence}.}} whose true community structures are known a-priori and whose properties 
 are summarized in Table~\ref{tab:dataset}.  \\ 
 {\bf Football} network was proposed by Girvan and Newman \cite{GN} which contains the network of American football games between Division
IA colleges during regular season Fall of 2000. The vertices in the graph represent teams (identified by their college names), and edges
represent regular-season games between the two teams they connect. \\
{\bf Indian Railway} network proposed by Ghosh et al.~\cite{Ghosh} consists of nodes representing stations, where two stations $s_i$ and
$s_j$ are connected by an edge if there exists at least one train-route such that both $s_i$ and $s_j$ are scheduled halts on that route. The
 weight of the edge between $s_i$ and $s_j$ is the number of train-routes on which both these stations are scheduled halts. We
 filter out the low-weight edges and then make the resultant network unweighted. We tag each station based on the
state in India\footnote{{\scriptsize\url{http://irfca.org/apps/station_codes}}} to which that station
belongs. The states
 act as communities since the number of trains within each state is much higher than the number of trains between two states. \\
{\bf Co-authorship} network is derived from the citation dataset\footnote{{\scriptsize\url{http://cnerg.org}}}
developed by
Chakraborty et al.~\cite{asonam}. This dataset contains the metadata (title, author(s), related field(s)\footnote{{\scriptsize Note that,
the different sub-branches like Algorithms, AI, Operating Systems etc. constitute the different ``fields'' of computer science domain.}} of the paper,
publication venue, year of publication, references and abstract) of all the papers of computer science published between 1960 to 2009
and archived in DBLP repository. We build an aggregated undirected coauthorship network where each node represents an
author, and an undirected edge between a pair of authors is drawn if they were co-authors at least once. Since each
paper
is marked by its related field, we assume this field as the research
area of the author(s) writing that paper. Therefore, an author may possess more than one area of research interests. We resolve this by
tagging each author by the major field on which she has written most of the papers. These fields act as the ground-truth communities. 
Besides the aggregated network, we also create some intermediate networks mentioned in Table~\ref{asymptotic} by cumulatively aggregating
all the vertices and edges over each year, e.g., 1960-1971, 1960-1972, ..., 1960-1980.   

Note that, the principles for constructing the Indian railway network and the co-authorship network are the same -- there is an underlying
bipartite structure in each case; for railway network, it is the station-train interaction network with an edge denoting if a particular
train passes through a station, while for the co-authorship network it is the article-author interaction network with an edge denoting the
authorship of a researcher in a scientific article. The railway network is therefore the one-mode projection of the train-station network
and the co-authorship network is similarly the one-mode projection of the article-author network. Note that, although the principles of
construction are same for both the networks (one clique per train/article is imposed in the one-mode projection), the results, as we shall
see in Section \ref{comm} are far better for the railway network since the ground-truth here is  much more fine-grained in comparison to
the co-authorship network.

\vspace{-5mm}
\begin{table}[!h]
\begin{center}
\caption{Properties of real-world networks; $n$ and $e$ are the number of nodes and edges, $c$ is the number of communities, <$k$> and
$k_{max}$ its average and maximum degree, $n_c^{min}$ and $n_c^{max}$ the sizes of its smallest and largest
communities.}\label{tab:dataset}
\vspace{2 mm}
 \scalebox{0.8}
 {
\begin{tabular}{l|c|c|c|c|c|c|c}
\hline
Networks            & $n$ & $e$ & <$k$> & $k_{max}$ & $c$ & $n_c^{min}$ & $n_c^{max}$ \\
\hline\hline
Football           & 115 & 613 & 10.57 & 12 & 12 & 5 & 13  \\ 
Railway           & 301 & 1224 & 6.36 & 48 & 21 & 1 & 46 \\
Coauthorship     & 103677 & 352183 & 5.53 & 1230 & 24 & 34 & 14404 \\

\hline
\end{tabular}}
\end{center}

\end{table}

\begin{table*}[!ht]
\caption{For football network, the values of the scoring functions on the output obtained from different algorithms and the
scores of the validation metrics with respect to the ground-truth communities. The ranks of the algorithms (using dense ranking) are shown
within parenthesis. The average ranks of all the normal (weighted) validation measures are
shown
in column 9 (column 13).}\label{table1}
\centering
 \scalebox{0.75}
 {
\begin{tabular}{|c|c|c|c|c||c|c|c|c||c|c|c|c|}
\hline
Algorithms & Mod & Perm & 1-Con & 1-Cut & NMI & ARI & PU & Avg & W-NMI & W-ARI & W-PU & Avg\\
     &     &    &    &     &      &     &    & (N) &        &      &      & (W)\\\hline     

Louvain	& 0.60(1) & 0.36(1) & 0.77(5) & 0.44(5) & 0.93(1) & 0.99(1) & 0.89(2) & 1.33 & 0.99(2) & 0.93(2) & 0.99(1)    & 1.67\\\hline
FastGreedy & 0.58(2) & 	0.25(3) & 0.81(3) & 0.59(3) & 0.93(1) & 0.99(1) & 0.91(1) & 1.00 & 1.00(1)& 0.94(1)&  0.99(1)&1.00\\\hline
CNM        & 0.55(3) & 0.20(4) & 0.85(1) & 0.86(1) &  0.67(4) & 0.75(4) & 0.42(5) & 4.33 & 0.55(5) &0.63(5) & 0.71(3) & 4.33\\\hline
WalkTrap   & 0.60(1) & 0.36(1) & 0.82(2) & 0.69(2) & 	0.90(2) & 0.98(2) & 0.84(3) & 2.33 & 0.98(3) & 0.91(3) & 0.99(1) & 2.33\\\hline
Infomod    & 0.60(1) & 	0.35(2) & 0.82(2) & 0.69(2) & 0.89(3) & 0.97(3) & 0.82(4) & 3.33 & 0.97(4) & 0.89(4) & 0.98(2) & 3.33\\\hline
Infomap    & 0.60(1) & 	0.35(2) & 0.79(4) & 0.51(4) &  0.89(3) & 0.97(3) & 0.82(4) & 3.33 & 0.97(4) & 0.89(4) & 0.98(2) & 3.33\\\hline

\end{tabular}}
\end{table*}

\begin{figure*}[!ht]
\centering
 \includegraphics[scale=0.35]{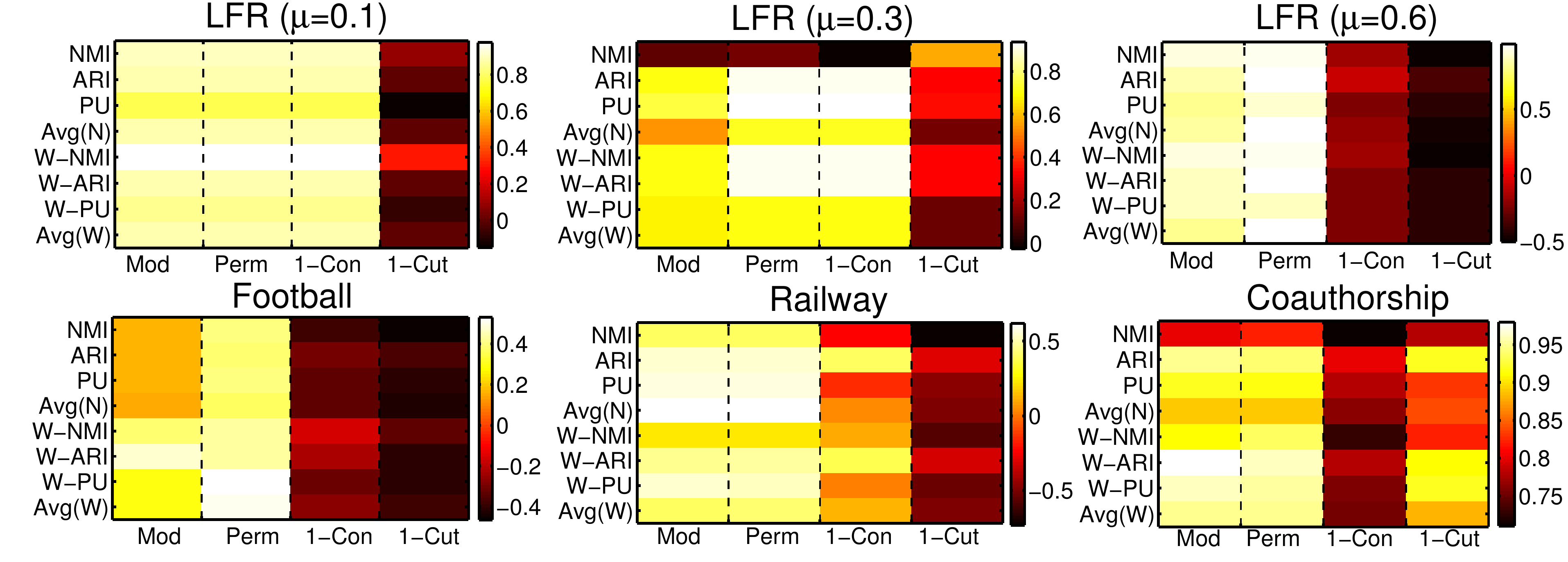}
 \caption{(Color online) Heat maps depicting the pairwise Spearman's rank correlation between four scoring functions with six validation
measures for six different networks. Avg(N) and Avg(W) are the averages of the ranks of three normal and three weighted validation measures
respectively as shown in Table~\ref{table1}.}\label{rank}
\end{figure*}

%

\vspace{-3mm}
\subsection{Scoring Functions for Evaluating Community Structure}\label{subsec:eval}
The goodness of a community is often measured by how well certain scoring functions are optimized. Here we compare the optimal value of
permanence for the obtained communities versus three popular scoring functions, namely modularity (Mod)~\cite{Newman:2006},
conductance (Con)~\cite{cond_09} and cut-ratio (Cut)~\cite{Leskovec:2010}. In order to make the higher the better, we measure (1-Con) and (1-Cut) for conductance and cut-ratio respectively. 


\subsection{Metrics to Compare with Ground-truth}\label{metrics}
A stronger test of the correctness of the community detection algorithm, however, is by comparing the obtained community with a given ground-truth structure.
We use three standard validation metrics, namely Normalized Mutual Information (NMI)~\cite{danon2005ccs}, Adjusted Rand Index (ARI)~\cite{hubert1985} and Purity (PU)~\cite{Manning} to measure the accuracy of the detected communities with respect to the ground-truth community structure. \cite{Labatut} argues that these measures have certain drawbacks in that they ignore the connectivity of the network. We therefore also use the weighted versions of these measures, namely Weighted-NMI (W-NMI), Weighted-ARI (W-ARI) and Weighted-Purity (W-PU) as proposed in~\cite{Labatut}. Note that, all the metrics are bounded between 0 (no matching) and 1 (perfect matching).


\subsection{Community Detection Algorithms}\label{algorithms}
We use the following community detection algorithms for comparison with our proposed algorithm discussed in
Section~\ref{comm}:\\
\textbf{(i) Modularity-based:}  FastGreedy \cite{newman03fast}, Louvain \cite{blondel2008} and CNM \cite{Clauset2004}.\\
\textbf{(ii) Random walk-based:} WalkTrap \cite{JGAA-124}. \\
\textbf{(iii) Compression-based:} InfoMod \cite{rosvall2007} and InfoMap~\cite{Rosvall29012008}.

\begin{figure*}
\centering
\includegraphics[scale=0.30]{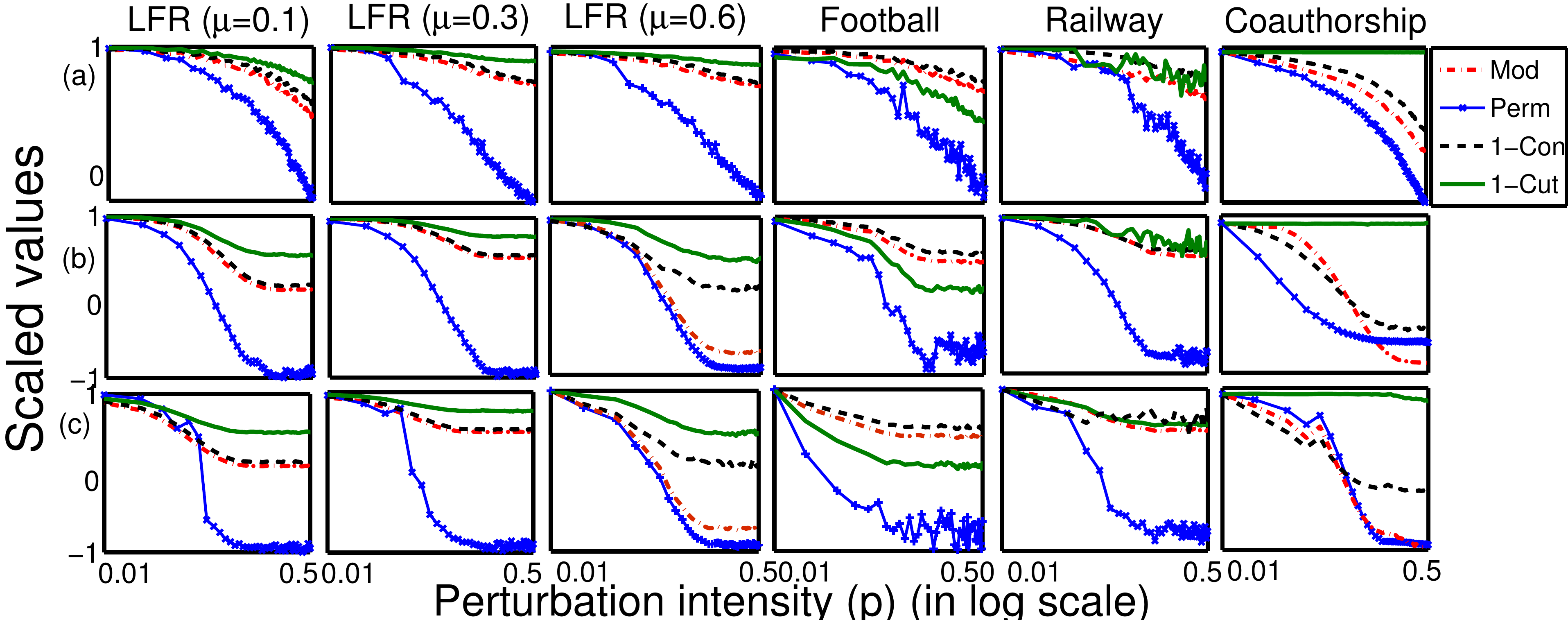}
\caption{(Color online) Change in the value of the scoring functions with the increase of perturbation intensity ($p$) in (a) edge-based,
(b) random and (c) community-based perturbation strategies. The values are normalized by the maximum value obtained from each function. }
\label{fig:perturbation}
\end{figure*} 

\section{Permanence as a Community\\ Scoring Function}\label{goodness}

In this section, we demonstrate the effectiveness of permanence as a scoring function for
evaluating the goodness of detected communities, and compare it with 
modularity, 1-Con and 1-Cut. To do this, we perform the following experiment, 
on the same lines as that of~\cite{steinhaeuser2010}.

These are the steps in our experiment: (i) We apply several community detection algorithms on a specified network and obtain the vertex-to-community assignment as given by  each algorithm; (ii) We compute the values of all the community scoring functions for these communities; (iii) For each scoring function we rank the algorithms based on which one of these produces the most optimal (highest) value; (iv) We then compare the obtained community with the known ground-truth community
and compute the respective validation measures, namely NMI, ARI, Purity and their weighted versions;  (v) For each validation metric, we rank the algorithms based on the one that produces the highest value, i.e., best match with ground-truth.  

Table~\ref{table1} shows the results of the experiment performed on football network. 
\if{0}
For the scoring function modularity, the rank of Louvain, WalkTrap, Infomod and Infomap are all set to 1, because they give the highest value. FastGreedy with the next highest value is ranked 2 and CNM with the lowest value is ranked 3. In the corresponding NMI column, Louvain and FastGreedy are ranked 1 because they have the highest value; WalkTrap is ranked 2 (second highest value); Infomod and Infomap are ranked 3 (third highest value) and CNM is ranked 4.
\fi
 Scoring functions (columns 2-5) are measures of goodness of the community set obtained. 
The validation metrics (columns 6-8, 10-12) measure the concurrence of the communities with the 
ground-truth communities.  We posit that since these two types of measures are orthogonal, and because the
validation metrics generally provide a stronger measure of correctness after 
measuring similarity with the ground-truth structure, the
values of a good scoring function should ``match" those of the validation metrics.
That is, if a scoring function indeed identifies the correct communities, then when its value is high (low), 
the values of the corresponding validation metrics would also be high (low).

To compute this correlation,
we compare the relative ranks, because the range of the values is not commensurate 
across the quantities and we are more interested in observing the ``up" or ``down" direction, rather than the absolute values.
For each network, we measure the Spearman's rank correlation between all pairs
of scoring functions and validation measures. Note that, it is not always possible to assign ranks uniquely. We used different ranking
schemes to break ties. Here, we present the results using dense ranking; we have also used
standard competition ranking and fractional ranking (see in \cite{appendix}) and our results are consistent
across all the different methods.

{\bf Results.} 
Table~\ref{table1} shows the values and ranks for the different metrics for football network. For all the networks, the rank
correlations of the scoring functions and the validation metrics are shown as heat maps in Figure~\ref{rank}. Lighter color
indicates higher correlation and hence more similarity between the scoring function and the validation metric.
For the networks having distinct community structure such as LFR ($\mu=0.1$), football and railway networks, permanence shows comparable performance as that of other scoring functions. However for LFR network, with the increase of $\mu$, the
inter-community connection density starts increasing, and it is difficult for any community detection algorithm and/or
 scoring function to capture the ground-truth communities. Interestingly, we observe that the rank correlation obtained through the permanence scores and those through validation metrics is exceptionally high for LFR ($\mu=0.6$) and coauthorship networks which seem to have poor community structure compared to the other networks (see Table~\ref{tab:dataset} and Table~\ref{similarity}). 
Since the ground-truth communities are not well formed, there is a wide variance in the type of community structures identified 
by different algorithms. Permanence score can capture this variability much better than other scoring functions. To summarize the results, in Table~\ref{avg} we present the average rank correlations of these community scoring functions across all the validation metrics for each network. We observe that for all the networks, permanence on an average produces the best ranking followed by modularity, conductance and cut-ratio in order. 
 
%

\vspace{-0.5cm}
\begin{table}[!ht]
\caption{Average rank correlations (over different validation metrics) of the community scoring functions  for all the networks. For each network, the highest (best) correlation is in bold font.}\label{avg}
\centering
 \scalebox{0.8}{
\begin{tabular}{|c|c|c|c|c|}
\hline
Networks & Modularity & Permanence & Conductance & Cut\\\hline
LFR($\mu$=0.1) & {\bf 0.88} & {\bf 0.88} & {\bf 0.88} & 0.02\\\hline
LFR($\mu$=0.3) & 0.61 &{\bf  0.74} & 0.72 & 0.28\\\hline
LFR($\mu$=0.6) & 0.87 & {\bf 0.96} & -0.18 & -0.44\\\hline
Football & 0.25 & {\bf 0.43} & -0.29 & -0.41 \\\hline
Railway & 0.43 & {\bf 0.46} & 0.08 & -0.48 \\\hline
Coauthorship & {\bf 0.92} & {\bf 0.92} & 0.76 & 0.86\\\hline
\end{tabular}}
\end{table}

\vspace{-3mm}

\section{Sensitivity of Permanence}\label{parturbation}

We now evaluate the sensitivity of permanence under different perturbations of the ground-truth community structure. We posit that a good metric for evaluating communities
 should be stable under small perturbations of the ground-truth communities (i.e., groups of nodes that differ
very slightly from the ground-truth communities). This indicates that the scoring function is robust to noise.
  However, if the perturbation is beyond a threshold, i.e., when 
the ground-truth community structure is perturbed to such an extent that
it resembles a random set of nodes, then a good scoring function should assign it a low score. 

Given a graph $G=<V,E>$  and  \emph{perturbation
intensity}  $p$, we  start with the ground-truth community $S$ and then modify it (i.e., change its members) by executing the perturbation strategy $p\cdot m$ times. The value of $m$ is based on different strategies, as described below. For our experiments, we adopt three perturbation strategies motivated by the methods proposed in~\cite{Yang:2012}:

 {\bf (i)Edge-based} perturbation picks a random inter-comm-unity edge $(u,v)$ where $u \in S$ and $v \in S'$ (where $S \neq S'$) and then swaps
the memberships (i.e., assign $u$ to $S'$ and $v$ to $S$). It continues until $p \cdot |E|$ iterations are
completed (here, $m=|E|$). This strategy preserves the size of $S$. However, if $v$ is not connected
to any other nodes in $S$ except $u$, then it makes $S$ disconnected.

{\bf (ii)  Random} perturbation takes community members and replaces them with random non-members. We pick two random nodes $u \in S$ and $v \in S' $ (where $S \neq S'$) and then swap their memberships. It  continues until $p \cdot |V|$ iterations are
completed (here, $m=|V|$).  Random perturbation maintains the size of $S$ but may disconnect $S$.
Generally, it degrades the quality of $S$ faster than edge-based strategy, since edge-based strategy only affects the ``fringe'' of the
community.

{\bf (iii) Community-based} perturbation adopts a similar mechanism as in the edge-based
strategy. However, it considers each community $S$ from the ground-truth community structure one by one and continues the perturbation until $p \cdot |S|$ constituent nodes of the community are swapped with the other non-constituent nodes (here, $m=|S|$). This process is repeated for
all the communities separately. This perturbation decreases the quality of the ground-truth communities the fastest  among the three since the number of swaps is much higher than the others.

We perturb different networks using these three perturbation strategies for values of $p$ ranging between 0.01 to 0.5.  We compute the
values of four community scoring functions, i.e., modularity, permanence, 1-Con and 1-Cut. For small values of $p$, small
change of the original value of the scoring function is
desirable since it indicates that the scoring function is robust to noise. For high perturbation intensities (i.e., for larger values of
$p$), the value should drop significantly since the communities become more random. 


%

{\bf Results.} Figure~\ref{fig:perturbation} shows the  results of our experiments. For a commensurate comparison, we rescale the values of each parameter by normalizing  with the maximum value obtained from that function. For all three strategies,
the values of the scoring functions tend to decrease with the increase of $p$, and the effect is most prominent in community-based strategy followed
by  random and edge-based strategies. For each network, once $p$ has reached a certain threshold, the decrease is much faster in permanence. 
On more careful inspection, we find that this happens because the internal structure of a community completely breaks down if 
perturbation is taken beyond a point and thus has an avalanche effect on the value of the clustering coefficient ($c_{in}(v)$ in Equation (\ref{perm})). This in turn quickly pulls the value of permanence down. 
Summarizing, the results  indicate that permanence is a better measure for distinguishing true communities from randomized sets of nodes than the other parameters. 

\if{0}
In order to further observe how perturbation affects each of the three major components of the permanence metric, namely the internal degree ($I(v)$), the maximum external connections ($E_{max}(v)$) and the internal clustering-coefficient ($c_{in}(v)$), we further measure the change of their individual values as a function of $p$. Figure~\ref{fig:perturbation_comp} shows the rate of these changes for random perturbation.  The most sensitive components are the internal degree and the average internal clustering-coefficient of vertices. These values tend to be comparatively stable for small perturbations, but degrades significantly as $p$ increases. 

\begin{figure}[!ht]
\centering
\includegraphics[scale=0.20]{factor.pdf}
\caption{(Color online) Change in the average values of internal degree ($I(v)$), maximum external connections ($E_{max}(v)$) and internal clustering-coefficient ($c_{in}(v)$) of vertices of two representative networks with the increase of perturbation intensity in random perturbation strategy. }
\label{fig:perturbation_comp}
\end{figure} 

\fi

\vspace{-3mm}

\section {Permanence Maximization}\label{comm}
Inspired by the effectiveness of permanence as a scoring function and its sensitivity to perturbations, we develop a
 community
detection algorithm called {\bf Max\_Permanence}\footnote{{\scriptsize The code is available at
\url{http://cnerg.org/permanence}.} }
(pseudocode in Algorithm~\ref{algo}) that identifies communities by maximizing permanence. 

Our algorithm is a  heuristic, that  strives to obtain a high  value of permanence. 
In this algorithm, we begin with initializing every vertex to a singleton community. 
A vertex is moved to a community only if this movement results in a net increase in the number of internal connections
and/or a net decrease in the number of external connections to any of the neighboring communities. If such a move is not possible, then
either the vertex remains as a singleton 
(such as in the case where moving to any one of the candidate communities will give equal permanence) or moves to the community where it is more tightly connected with its neighbors (this causes the vertex to have positive permanence).  
This process is repeated for each vertex and the entire relocation of all vertices is repeated over several iterations until the permanence value converges.
However, convergence is not theoretically guaranteed, but we observed that the algorithm converges with high probability.

\begin{algorithm}[!t]
\scriptsize
\caption{ Max\_Permanence}\label{algo}
{\bf Input:}  A graph $G$.\\
  {\bf Output:} Permanence of $G$; Detected communities\\

 \begin{algorithmic}
   \Procedure{Max Permanence}{$G(V,E)$}
 \State Each vertex is assigned to its seed community
 \State Set value of maximum iteration as $maxIt$ 
 \State $vertices \gets |V|$
 \State $Sum \gets  0 $
  \State $Old\_Sum \gets  -1 $
  \State $Itern \gets 0 $
   \While{$Sum \neq Old\_Sum$ and $Itern < maxIt$}
    \State $Itern \gets Itern+1$
    \State $Old\_Sum \gets  Sum $
    \State $Sum \gets  0 $
  \ForAll {$v \in V $}\\
  \hspace{1.2 cm}(compute current permanence of $v$)
   \State $cur\_p \gets Perm(v)$
   \If {$cur\_p ==1$}
   \State $Sum \gets  Sum+cur\_p $\\
   \hspace{1.6 cm} {\bf continue};\\
   \EndIf
   \State $cur\_p\_neig \gets 0$
   \Comment $Neig(v)$=set of neighbors of $v$
   \ForAll{$u \in Neig(v)$} \\
     \hspace{1.6 cm}(compute current permanence of $u$)
   \State $cur\_p\_neig \gets cur\_p\_neig + Perm(u)$ 
   \EndFor\\\\
   \Comment $Comm(v)$ is the set of neighboring communities of $v$
   \ForAll {$C \in Comm(v)$}\
   \State Move $v$ to community $C$\\
   \hspace{1.7 cm}(compute permanence of $v$ in community $C$)
 \State $n\_p \gets Perm(v)$ \\\\
  \Comment Neighbors of $v$ are affected for this movement
  \State $n\_p\_neig \gets 0$
  \ForAll{$u \in Neig(v)$} \\
     \hspace{1.6 cm}(compute new permanence of $u$)
   \State $n\_p\_neig \gets n\_p\_neig + Perm(u)$ 
   \EndFor
 \If {($cur\_p < n\_p)$ and ($cur\_p\_neig< n\_p\_neig)$}
 \State $cur\_p \gets n\_p$
 \Else \\
      \hspace{2 cm} replace $v$ to its original community
 \EndIf
  \EndFor
   \State $Sum \gets  Sum+cur\_p $
   \EndFor
 \EndWhile
 \State $Netw\_perm=Sum/vertices$  \Comment Permanence of $G$
 \State \Return $Netw\_perm$
 \EndProcedure
 \end{algorithmic}
 \end{algorithm}

 \begin{table*}[!ht]
\caption{Improvement of Max\_Permanence with respect to the average (left-hand value) and the best (right-hand value) performances of the six competing algorithms (separated by semicolon). Positive (negative) values indicate that Max\_Permanence outperforms (underperforms) the corresponding performances of the competing algorithms.}\label{avg_Improvement}
\centering
\vspace{2mm}
 \scalebox{0.90}{
\begin{tabular}{|c|c|c|c|c|c|c|}
\hline
Validation metrics & LFR ($\mu$=0.1) & LFR ($\mu$=0.3) & LFR ($\mu$=0.6) & Football & Railway & Coauthorship \\\hline

NMI & 0.04;\ \  0.00 & 0.15;\ \ 0.05  & -0.31;\ \ -0.78 & 0.04;\ \ 0.00 & 0.15;\ \ 0.11 & 0.04;\ \ -0.06\\\hline 

ARI & 0.06;\ \ 0.00 & 0.21;\ \  0.02 & -0.39;\ \  -0.76  & 0.07;\ \  0.00 & 0.03;\ \  0.04 & 0.03;\ \  -0.08\\\hline

PU  & 0.04;\ \ 0.00 & 0.17;\ \  0.00 & -0.38;\ \  -0.72 & 0.01;\ \  0.00 & 0.13;\ \  0.00 & 0.03;\ \  -0.06 \\\hline

W-NMI & 0.02;\ \ 0.00 & 0.14;\ \  0.00 & -0.41;\ \  -0.78 & 0.09;\ \  0.00 & 0.26;\ \  0.00 & 0.05;\ \  -0.01\\\hline

W-ARI & 0.05;\ \ 0.02 & 0.19;\ \  0.05 & -0.35;\ \  -0.72 & 0.05;\ \  0.00 & 0.02;\ \  -0.15 & 0.04;\ \  -0.06 \\\hline

W-PU & 0.03;\ \ 0.01 & 0.17;\ \  0.00 & -0.45;\ \  -0.79 & 0.00;\ \  0.00 & 0.05;\ \  -0.04 &  0.02;\ \  -0.15\\\hline            
                 
 \end{tabular}}
\end{table*}

\subsection{Performance Evaluation}
Table~\ref{avg_Improvement} shows results of the improvement  of our method (as differences) compared 
to the average and best performances of six competing algorithms (given in Section~\ref{algorithms})  based on six ground-truth based
validation
metrics. 

{\em Comparable results } - in LFR ($\mu$ = 0.1) and football networks, since the communities are well-separated, most algorithms efficiently capture these partitions and our method is comparable to the other algorithms as well. 

{\em Improved results } - in LFR ($\mu=0.3$) and railway networks, our method significantly outperforms other algorithms. 
Moreover in railway network, we observe that
our algorithm detects three singleton communities (i.e., communities each containing only one
vertex), one of which is also present in 
 the ground-truth structure. None of the community detection algorithms is able to capture this singleton community. 

{\em Moderate results} - our method does not work well for the LFR ($\mu=0.6$) network. 
For coauthorship network, we observe that though our algorithm outperforms the average performance of the competing algorithms, it performs less well than that of the two information-theoretic approaches (Infomod and Infomap).\\ 

\noindent{\bf Reasons behind moderate performance }\\
 \underline{LFR ($\mu$=0.6)} --  
 To understand why our algorithm is not as competitive for
 LFR ($\mu=0.6$), we further observe the quality of the ground-truth communities in three LFR networks. We observe that while the average internal
clustering coefficient of vertices in LFR ($\mu=0.1$) is 0.78, it deteriorates to 0.36 for LFR
($\mu=0.6$). Moreover, 97\% of vertices in ground-truth communities of LFR ($\mu$=0.6) have less internal connections than the
external connections (while LFR ($\mu$=0.1) and LFR ($\mu$=0.3) hardly have any such nodes). 
This indicates that LFR ($\mu=0.6$) does not have modular structure in the ground-truth communities. 

To further strengthen this claim, we also
measure the similarity of the
communities obtained by different community detection algorithms (as listed in Section \ref{algorithms})  across different validation
measures.  
The results in Table~\ref{similarity} clearly show the degradation of the similarity values with the increase in $\mu$. The
reason is that with the increase in $\mu$, the communities in LFR network tend to be less well-knit, and thus the agreement of the outputs
of different algorithms is also less.
Therefore, the output
of a good community detection algorithm should reflect such absence of modular structure in the network (hence shows poor performance).\\
\underline{Coauthorship network} -- To explain the permanence-based results obtained from coauthorship network, we further analyze
the communities obtained from our algorithm. We check the title and the abstract of the papers written by the authors in each
community of coauthorship network, and notice that our method splits large ground-truth communities into denser submodules. This
splitting is mostly noticed in older research areas such as Algorithms and Theory, Databases etc. These submodules are actually the
subfields (sub-communities) of a field (community) in computer science domain. Few examples of such sub-communities obtained from our
algorithm are noted in Table~\ref{subfield}.
Thus, our algorithm, in addition to identifying well-defined communities, is also able to unfold the hierarchical organization of a network.

\vspace{-5mm}
\begin{table}[!ht]
\caption{Average values among pairwise similarities between outputs of the community detection algorithms on different LFR networks.}\label{similarity}
\centering
 \scalebox{0.75}
 {
\begin{tabular}{|c|c|c|c|}
\hline
Validation & LFR  & LFR  & LFR \\
measures  &  ($\mu$=0.1)                &      ($\mu$=0.3)            &       ($\mu$=0.6)        \\\hline
NMI       & 0.95 & 0.82 & 0.53 \\\hline
ARI   &  0.98 & 0.79 & 0.48 \\\hline
PU   & 0.99  & 0.85 & 0.56 \\\hline
W-NMI & 0.94 & 0.85 & 0.54 \\\hline
W-ARI  & 0.97  & 0.78 & 0.50\\\hline
W-PU   & 0.98 & 0.83 & 0.57 \\\hline

\end{tabular}}
\end{table}

\vspace{-7mm}
\begin{table}[!ht]
\caption{Example of communities and sub-communities obtained from coauthorship network using Max\_Permanence algorithm. }\label{subfield}
\centering
 \scalebox{0.76}
 {
\begin{tabular}{|c|c|}
\hline
Communities & Sub-communities\\\hline
Algorithms & Theory of computation; Formal methods; Data structure; \\
and Theory & Information \& coding theory; Computational geometry\\\hline
Databases & Models; Query optimization; Database languages;\\
          & storage; Performance, security, and availability\\\hline
\end{tabular}}
\end{table}

\noindent {\bf Comparison of largest community size.} 
Many optimization algorithms have the tendency to underestimate smaller size communities (known as the resolution limit problem \cite{good2010}) and sometimes
tend to produce very large size communities. In our test suite, we observe the similar tendency in all the competing algorithms whereas the communities obtained by permanence are smaller in size.  In Table~\ref{max_comm}, we show for two representative networks that the size of the largest communities detected by the other algorithms is much larger than
 the size of the largest community present in the ground-truth structure. We also measure the maximum similarity (using Jaccard
coefficient)
between the largest-size community detected by each algorithm with the communities in ground-truth structure and notice that Max\_Permanence
is able to detect largest size community which is most similar to the ground-truth structure. Therefore, we hypothesize that our algorithm has the potentiality to reduce the effect of resolution limit. 

\vspace{-4mm}
\begin{table}[!ht]
\caption{Size of the largest communities obtained from different community detection algorithms and their similarities with the ground-truth
structure for two networks (LFR ($\mu$=0.3, N=1000) and football).}\label{max_comm}
\centering
 \scalebox{0.70}
 {
\begin{tabular}{|c|c|c|c|c|}
\hline
 & \multicolumn{2}{c|}{Largest community size}& \multicolumn{2}{c|}{Similarity} \\\cline{2-5}
 &  LFR  & Football & LFR  & Football \\
 & ($\mu=0.3$)  & & ($\mu=0.3$) & \\\hline
Ground-truth & 49 & 12 & -- & -- \\\hline
Louvain & 62 & 24 & 0.70 & 0.41 \\\hline
FastGreedy & 95 & 18 & 0.32 & 0.65   \\\hline                                    
CNM  & 91 & 32 & 0.52 & 0.31 \\\hline
Walktrap  & 83 & 15 &  0.51 & 0.57 \\\hline
Infomod & 61 & 16  &  0.79 &   0.86 \\\hline
Infomap & 59 & 16  &  0.74 &   0.86\\\hline
Max\_Permanence & 49 &   13 &   1 &  0.92  \\\hline

\end{tabular}}
\end{table}

%


  
%
%
%
%
%
%
%
%
%

\vspace{3mm}
\subsection{Handling Modularity Maximization Issues}\label{permanence_property}
\label{perm_property}
As discussed earlier, modularity maximization algorithms suffer from the issues including (a) resolution limit, (b) degeneracy of solution and (c) dependence on the size of the graph~\cite{good2010}. We now discuss how each of these problems are ameliorated by maximizing permanence.

%

\begin{figure}[!h]
\centering
\psfrag{al}{$\alpha$}
\psfrag{be}{$\beta$}
\includegraphics[scale=0.3]{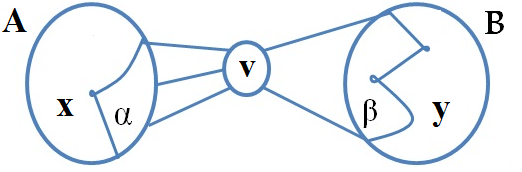} 

 \caption{(Color online) An illustrative example to show the community assignment of vertex $v$.}\label{example1}
\end{figure}

We demonstrate that community assignments are different in a modularity-based algorithm vis-a-vis {\bf Max\_Permanence} algorithm
using the example in Figure~\ref{example1}. In this figure, we assume that apart from the edges
through $v$, there is no connection between the communities $A$ and $B$. 

\begin{table*}[!ht]
\caption{Change in scoring functions with the (near-)symmetric growth of coauthorship network obtained in each year by adding vertices and
edges till that year. $N$:
number of nodes, $c$: number of communities, $<I>$: average internal degree, $<k>$: average degree, $<c_{in}>$: average internal clustering
coefficient, $<E_{max}>$: average maximum external connectivity. The value of permanence is
less affected by the growth. 
 }\label{asymptotic}
\centering
 \scalebox{0.8}
 {
\begin{tabular}{|c|c|c|c|c|c|c|c|c|c|c|c|c|}
\hline
\multirow{9}{*}{\begin{sideways}Coauthorship\end{sideways}} & \multicolumn{2}{c|}{Year} & 60-71 & 60-72 & 60-73& 60-74& 60-75 &
60-76 & 60-77 & 60-78 & 60-79 & 60-80 \\\cline{2-13}
 & \multirow{5}{*}{Network } &  $n$ & 964 & 1515 & 1991 & 2681 & 3386 & 4836 & 6284 & 7814 & 9001 & 10386 \\\cline{3-13}
			      & \multirow{5}{*}{properties} & $c$ & 24 & 24 & 24 & 24 &24 &24 &24 &24 &24 &24 \\\cline{3-13}
			      & &  $\frac{<I>}{<k>}$ &  0.082 & 0.095 & 0.093 & 0.091 & 0.089 & 0.104 & 0.111 & 0.112 & 0.115 & 0.113 \\\cline{3-13}
			      & &   $\frac{1}{<E_{max}>}$ ($\times 10^{-4}$) & 3.8 & 3.2 & 2.9 & 3.9 & 2.8 & 2.11 & 2.39 & 2.92 & 2.69 & 3.22\\\cline{3-13}
			      & & $<(1-c_{in})>$ & 0.239 & 0.248 & 0.246 & 0.251 & 0.251 & 0.260 & 0.265 & 0.269 & 0.270 &
0.274\\\cline{2-13}
			      & \multicolumn{2}{c|}{Modularity} & 0.369 & 0.374 & 0.395 & 0.392 & 0.421 & 0.422 & 0.465 & 0.471 & 0.493 & 0.501\\\cline{2-13}
			      & \multicolumn{2}{c|}{Permanence} & 0.094 & 0.092 & 0.092 & 0.096 & 0.095 & 0.095 & 0.097 & 0.097 & 0.097 & 0.098 \\\hline

\end{tabular}}
\end{table*}

{\bf Terminology.}
Let vertex $v$ be connected to $\alpha$ ($\beta$) nodes in community $A$ ($B$), and these $\alpha$ ($\beta$) nodes form the set $N_\alpha$ ($N_\beta$). The number of vertices in community $A$ is ($x+ \alpha$), and in community $B$ is ($y + \beta$).
 Let the average internal degree of a vertex $a\in N_\alpha$ and a vertex $b\in N_\beta$, before $v$ is
assigned to any of the communities, be $I_{\alpha}$ and $I_{\beta}$ respectively. Let the average internal clustering coefficient of the neighboring nodes in communities $A$ and $B$ be $C_A$ and $C_B$ respectively. If $v$ is added to
communities $A$ ($B$) then the average internal 
 clustering coefficient of $v$ becomes $C^{v}_{A}$ ($C^{v}_{B}$), and the average internal clustering coefficient of the nodes in $N_\alpha$($N_\beta$) become $C^{\alpha}$ ($C^{\beta}$). 
 
 We assume that the communities $A$ and $B$ are tightly connected internally such that the values of $C_A$ and $C_B$ are very high (> 0.5).
To simplify the explanations, we consider the case where none of the neighbors of $v$ are connected to each other. If $v$ does not add any
new edges to the group of neighbors, then  $C^{\alpha}=C_A\frac{(I_{\alpha}-1)}{(I_{\alpha}+1)}$
(similarly, $C^{\beta}=C_B\frac{(I_{\beta}-1)}{(I_{\beta}+1)}$). 

 
 Given this scenario, we can determine the conditions (due to the lack of space detailed calculations are provided in an online appendix \cite{appendix}) for which a particular assignment of $v$ to any
of the communities will give the highest permanence. Using these conditions, we show how permanence overcomes some of the issues related to 
modularity maximization.
 
 \noindent {\bf {Degeneracy of solution}} is a problem where a community scoring function (e.g., modularity) admits multiple distinct high-scoring
solutions and typically lacks a clear global maximum, thereby, resorting to tie-breaking~\cite{good2010}.  For our example, when 
$\alpha$ = $\beta$, modularity maximization algorithm will assign $v$ arbitrarily to $A$ or $B$. However, in the case of permanence,  $v$ will remain as a separate community so long as the following condition is maintained:

{\em Condition.
If $\alpha=\beta$, $C^{\beta}= C_B\frac{I_{\beta}-1}{I_{\beta}+1}$, then communities $A$, $B$ and $v$ will remain separate rather than $v$ joining community $A$, if 
 $\alpha(\frac{2C_A-1}{I_{\alpha}+1}) +(1-C^v_A) \ge \frac{1}{2\alpha}$.}

We observe that when $\alpha=\beta=1$, then $C^v_A=0$ and the communities will always remain separate. Furthermore, as $\alpha$ increases,
the left-hand side of the above condition will become larger than the right, thus increasing the chance of separate communities. 

 \noindent {\bf {Resolution limit}} is a problem where communities of certain small size are merged into larger ones~\cite{good2010}. A classic example where modularity cannot identify communities of small size is a cycle of $m$ cliques. 
 Here maximum modularity is obtained if two neighboring cliques are merged. 
 
 In the case of permanence, we can determine that whether two communities $A$ and $B$ would merge (as in modularity) or whether $v$ would join community $A$ (we select $A$, but similar analysis can also be done for the case when $v$ joins $B$), by the following condition:
 
{\em Condition. 
Joining $v$ to community $A$  gives higher permanence than merging the communities $A$, $B$ and $v$ if $C^{\beta}= C_B$, and
($\frac{\gamma}{(\gamma+1)\beta} +\frac{C^v_A(2\gamma+1)-C^v_B}{(\gamma+1)^2} -\frac{\beta}{I_{\beta}+1}$)>1;  where $\gamma = \alpha/\beta$
and also if $C^{\beta}= C_B\frac{I_{\beta}-1}{I_{\beta}+1}$, and $(\frac{\gamma}{(\gamma+1)\beta}
+\frac{C^v_A(2\gamma+1)-C^v_B}{(\gamma+1)^2}
+\frac{\beta(2C_B-1)}{I_{\beta}+1})>1$.  }
  
 
 This result is independent of the size of the communities. Moreover, so long as $A$ and $B$ are almost cliques (internal clustering
coefficients > 0.5), $C^v_A$ is sufficiently high and $C^v_B$ is sufficiently small (e.g., $C^v_A$ >2/3 and $C^v_B$=0), $v$ will join
community $A$ rather than merging. Thus, in general, {\em the highest permanence is obtained if $v$ joins the community to which it is very
tightly connected rather than the one to which it is loosely connected.}
 
\noindent {\bf Asymptotic growth of value} of a metric implies a strong dependence on the size of the network and the number of modules the
network contains~\cite{good2010}. Rewriting Equation~\ref{perm}, we get the permanence of the entire network $G$ as follows:
{\scriptsize$Perm(G)=\frac{1}{|V|} \sum_{v \in V}\left[\frac{I(v)}{D(v)E_{max}(v)}\right]-\frac{1}{|V|}\sum_{v \in
V}\left[(1-c_{in}(v))\right]$}. We can notice that most of the parameters in the above formula are independent of the network size and the
number of communities. Table~\ref{asymptotic} illustrates the change in modularity and permanence with the symmetric growth of
the network size in coauthorship network. Note that, the intermediate networks are formed by cumulatively
aggregating all the vertices and edges of coauthorship network over the years, e.g., 1960-1971, 1960-1972,..., 1960-1980. We observe
that the modularity increases consistently with the symmetric growth, while the value of permanence remains almost constant.

\if{0}
\subsection {Scenario and Parameters}
We assume a more generalized version of this problem where the clustering coefficients $C_A$, $C_B$, $C^v_A$ are high (greater than 1/2) and $\beta$=1. Because $\beta=1$, therefore $C^v_B=0$ and $\frac{\beta}{I_{\beta}+1}$ will be low.  In this case, $v$ will join community $A$ rather than community $B$. Furthermore, by Corollary ~\ref{c1}, $v$ will join community $A$ rather than merging $A$, $B$ and $v$.

We now describe with the illustrative example in Figure~\ref{example} (b) that which of these four assignments produce high permanence value for the entire graph. In Figure~\ref{example} (b), $x$=1, $y$=2, $\alpha=4$, $\beta=3$, $I_{\alpha}=\frac{13}{4}$, $I_{\beta}=\frac{7}{3}$, $C^{v}_{A}=\frac{5}{6}$, $C^{v}_{B}=\frac{2}{3}$, $C^{\alpha}=\frac{2}{3}$, $C^{\beta}=\frac{2}{3}$, $C_A=\frac{2}{3}$ and  $C_B=\frac{2}{3}$

These assignments will be determined based on the connections of $v$ to communities $A$ and $B$, the internal degrees of the vertices in
$N_A$ and $N_B$ and the internal clustering coefficients. In this paper, we highlight some of the special cases, which will help demonstrate
how permanence improves on the resolution limit and reduces degeneracy.  We do not present the proofs here due to limitation of space.
However, the proofs can be found at {\url {http://cse.iitkgp.ac.in/resgrp/cnerg/appendix_permanence}}.

\begin{lemma} \label{l1}
Given $C^{\alpha}=C_A$ and $C^{\beta}=C_B$, let $Z=\frac{\alpha -\beta}{\alpha \beta} +\left ( C^{v}_A -C^{v}_B \right)+\left( \frac{\alpha}{I_{\alpha}+1} - \frac{\beta}{I_{\beta}+1}  \right ) $. 
The assignment $[(A+v):B]$ will have a higher permanence than $[A:(v+B)]$, if $Z>0$  and a lower permanence if $Z<0$.
\end{lemma}

 \begin{lemma}\label{l2}
Merging the communities $A$, $B$ and $v$, gives higher permanence than joining $v$ to community $A$  if \\  $C^{\beta}= C_B$, and
$\frac{\gamma}{(\gamma+1)\beta} +\frac{C^v_A(2\gamma+1)-C^v_B}{(\gamma+1)^2} -\frac{\beta}{I_{\beta}+1}<1$;  where $\gamma = \alpha/\beta$. 
 and also if; \\ $C^{\beta}= C_B\frac{I_{\beta}-1}{I_{\beta}+1}$, and $\frac{\gamma}{(\gamma+1)\beta} +\frac{C^v_A(2\gamma+1)-C^v_B}{(\gamma+1)^2}
+\frac{\beta(2C_B-1)}{I_{\beta}+1}<1$.  
   \end{lemma}

\begin{lemma}\label{l3}
If  $C^{\alpha}= C_A$ and $C^{\beta}= C_B$  then the 
communities will merge  ($[(A+v+B)]$), rather than remain separate.\\  If  $C^{\alpha}= C_A\frac{(I_{\alpha}-1)}{(I_{\alpha}+1)}$ $C^{\beta}=
C_B\frac{(I_{\beta}-1)}{(I_{\beta}+1)}$and  then the communities will merge if:\\
$\frac{\gamma^2 C^v_A+C^v_B}{(\gamma+1)^2}  >\alpha\frac{(2C_A-1)}{I_{\alpha}+1} +\beta\frac{(2C_B-1)}{I_{\beta}+1}$.
\end{lemma}

\begin{lemma}\label{l4}
If  $C^{\alpha}= C_A$ and $C^{\beta}= C_B$  then the
communities will remain separate ($[A:v:B]$) rather than $v$ joining with community $A$ ($[(A+v):B]$), if;\\
$\alpha(\frac{1}{I_{\alpha}+1}+\frac{1}{(\alpha+\beta)\beta}) < (1-C^v_A)$ \\
Otherwise; If  $C^{\alpha}= C_A\frac{(I_{\alpha}-1)}{(I_{\alpha}+1)}$ $C^{\beta}= C_B\frac{(I_{\beta}-1)}{(I_{\beta}+1)}$and  then the communities will remain separate if;
$\alpha(\frac{2C_A-1}{I_{\alpha}+1}) +(1-C^v_A) \ge \frac{\alpha}{(\alpha+\beta)\beta}$
\end{lemma}

{\bf TODO This paragraph has to be written with the new figure}
As we can see from the equations in these Lemmas, the assignments are determined by the relative strength of the connection of $v$ to the two communities ($\alpha$ with respect to $\beta$ and  $C^v_A$ with respect to $C^v_B$). The four lemmas together can help determine the best assignment. In the example Figure; we see that by Lemma 1;... by Lemma 2...
Also on comparing the permanence values we see that this indeed is the case.

We will now use these Lemmas  to demonstrate how permanence as compared to modularity can overcome the resolution limit, reduce both the degeneracy of solutions and asymptotic growth of value with the increase of the network size and number of communities.

 \begin{corollary}\label{c1}
  If $\beta=1$, $C^{\beta}= C_B\frac{I_{\beta}-1}{I_{\beta}+1}$,  $C^v_A >1/2$  then  $v$ will join community $A$ rather than the three pieces merging.
  \end{corollary}
  
  \begin{corollary}\label{c2}
  If $C_B \approx 1$, $C^{\beta}= C_B\frac{I_{\beta}-1}{I_{\beta}+1}$, $\beta \ge I_{\beta}+1$ and  $C^v_A \ge C^v_B/3$  then  $v$
will join community $A$ rather than the three pieces merging.
  \end{corollary}

If $A$ and $B$ are almost cliques, then we can further increase the size of $\beta$, so long as $\beta \ge I_{\beta}+1$. Then $C^v_A=C^v_B$. If $\alpha$ is large enough that $\frac{\alpha}{I_{\alpha}+1} > \frac{\beta}{I_{\beta}+1}$ and $\alpha > \beta$, then too, by Corollary~\ref{c2} $v$ will join community $A$ rather than merging $A$, $B$ and $v$.

 \begin{corollary}\label{c3}
    If $\alpha=\beta$, $C^{\beta}= C_B\frac{I_{\beta}-1}{I_{\beta}+1}$,  $C^v_A =C^v_B$  then  communities $A$, $B$ and $v$ will merge, rather than
$v$ joining community $A$, if  $\frac{1}{2\beta}+\frac{C^v_A}{2}+\beta\frac{2C_B-1}{I_{\beta}+1}<1$.
    \end{corollary}

 \begin{corollary}\label{c4}    
 If $\alpha=\beta$, $C^{\beta}= C_B\frac{I_{\beta}-1}{I_{\beta}+1}$, then communities $A$, $B$ and $v$ will remain separate rather than $v$ joining community $A$, if 
 $\alpha(\frac{2C_A-1}{I_{\alpha}+1}) +(1-C^v_A) \ge \frac{1}{2\alpha}$.\\
 If $\alpha=\beta=1$, then $C^v_A=0$  and then the communities will always remain separated.
 \end{corollary}

\fi

\vspace{-0.2cm}
\section{Related Work}\label{related_work}

A huge volume of work has been devoted to finding communities in large networks, including diverse methods such as modularity optimization
\cite{blondel2008, Clauset2004}, spectral
graph-partitioning \cite{Newman_13}, random-walk \cite{JGAA-124}, information-theoretic
\cite{rosvall2007, Rosvall29012008}, consensus clustering \cite{Santo} and many others
(see~\cite{Fortunato} for the review). 
Recently, Chakraborty et al.~\cite{chakraborty} pointed out how vertex ordering influences the results of the community
detection algorithms. They identify invariant groups of vertices (named as ``constant communities'') whose assignment to communities are not
affected by vertex ordering.

On the other hand, several metrics for evaluating the quality of community structure
have been proposed. The most popular is modularity~\cite{Newman:2006}. 
However,
community detection using modularity has certain issues including resolution limit, degeneracy of solutions and asymptotic growth
\cite{good2010}. To address these issues, multi-resolution versions
of modularity \cite{Arenas} were proposed to allow researchers to specify a tunable target resolution limit
parameter. 
Furthermore, Lancichinetti and Fortunato~\cite{santo_11} stated that even those multi-resolution versions of modularity are not only inclined to merge
the smallest well-formed communities, but also to split the largest well-formed communities.


\section{Discussions and Future work}\label{discussion}

In this paper, we have introduced a new vertex-based metric, permanence for evaluating the goodness of communities in networks.  From our experiments we observe that the  permanence score has a good correlation with the quality of the ground-truth communities (Section~\ref{goodness})  and is sensitive to perturbations in  the community structure (Section~\ref{parturbation}). In addition, permanence also provides some significant advantages compared to other popular community scoring functions. 

The {\em value of permanence strongly correlates to the community like structure of the network}.  For example, 
the power grid network, which is not at all modular~\cite{Karrer}, has a modularity of 0.98  and a permanence of -0.16. In
contrast, community-rich networks such as a circle of 30 cliques 
 generate permanence (modularity) of 0.92 (0.87). Therefore, permanence can also be used
 to identify whether the network is at all suitable for community
detection. 
\if{0}
 Figure~\ref{perm_dist} depicts the cumulative distributions of vertex-level permanence 
for the networks in our test suite.
The networks with well-defined community structure show flat tail at the end with some vertices having the highest permanence (=1). But for non-modular 
networks (e.g., LFR ($\mu=0.6$)), the end-slope is quite sharp and finishes near at 0.2-0.3. For networks like LFR ($\mu=0.3$) and Football, the curves lie in-between these two extreme cases. Railway and coauthorship networks contain  few vertices that are very stable (high permanence)  in their assigned communities, while a large number of vertices seem to be unstable. The 
vertices with high (low)-permanence in railway network are highly (loosely)-connected older (newer) stations. In coauthorship network, the high (low)-permanence nodes are the experienced (novice) authors with huge (almost no) collaborations. 

\begin{figure}[!h]
\begin{center}
\includegraphics[scale=0.18]{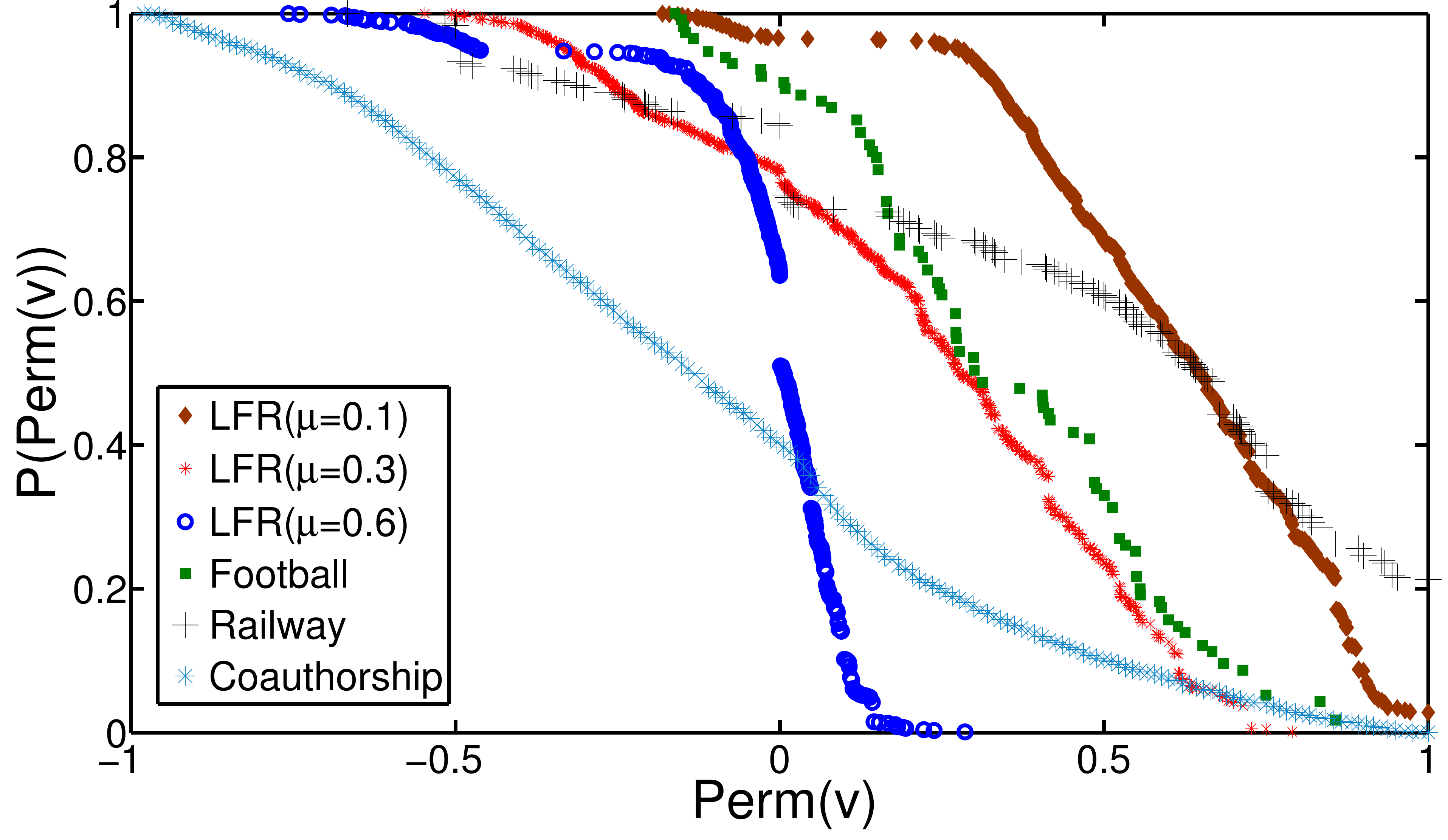}
\caption{(color online) Cumulative distribution of vertex-level permanence in different networks.}\label{perm_dist}
\end{center}
\end{figure}

Because the value of permanence is influenced more by the community structure of the network, it is {\em less susceptible to  issues common with modularity maximization} such as resolution limit and asymptotic growth of value with network size. As discussed in Section~\ref{comm}, the largest community obtained by permanence, although smaller than the competing methods, is yet the one that most matches the ground-truth. This result clearly indicates that permanence tries to find the exactly correct size community, which is the key to resolving resolution limit. 
We get further confirmation from the fact that permanence  allows us to identify singleton communities (that act as a bridge between two communities) as in
LFR ($\mu$ = 0.6) and railway networks, as well as a few small-size communities in the railway and coauthorship networks. The
small communities
in railway network are the highly-connected junctions (sub-communities) within a state (communities); those in coauthorship networks are different subfields (sub-communities) within a field (community). We have also shown in Table~\ref{change}, that compared to modularity, the value of permanence remains more stable with the growth in the network size. 
\fi

We believe that the advantages of permanence arise because it is a local vertex-based metric as opposed to the more common global/mesoscopic
metrics. At the same time, permanence also derives the benefits of a global metric to a certain extent by looking into the exact community
assignments of the external neighbors of the vertex considered. Perfectly global metrics tend to aggregate the effect of the connections of
all the vertices in a community.  As we have seen in Section~\ref{permanence_def} we can lose information by aggregation, particularly if
the distribution of the connections is skewed.  A vertex-based metric is more fine-grained, and therefore allows partial estimation of 
communities in a network whose entire structure is not known. 


In this paper we have empirically demonstrated the 
advantages of permanence. As an immediate future work, we plan to extend
permanence metric to evaluate
the quality of overlapping communities and communities in dynamic and weighted networks. We believe that
this metric will help in formulating a strong theoretical foundation for identifying
community structures where the ground-truth is not known. All the codes, datasets and supporting materials are publicly
available at \url{http://cnerg.org/permanence/}.



\if{0}
\todo{Tanmoy: Local vs Global}
\todo{Tanmoy: Resolution limit}
\todo{Tanmoy: Move to the discussion}
\todo{TANMOY: small survey on local measures}
\todo{Isn't there other local measures -- then why this new}
\todo{Tanmoy please add few references for local measures.}
\fi




\vspace{-2mm}
\section{Acknowledgments}
The first author of the paper is financially supported by Google India PhD Fellowship Grant for Social Computing. The authors from
University of Nebraska are supported by College of IS\&T, the Graca grant for UNO-SPR and the RISC Fellowship.

%

\vspace{-2mm}
\bibliographystyle{abbrv}
\bibliography{bib1}

\end{document}